\documentclass{article}

\usepackage{arxiv}

\usepackage[utf8]{inputenc} 
\usepackage[T1]{fontenc}    
\usepackage{hyperref}       
\usepackage{url}            
\usepackage{booktabs}       
\usepackage{amsfonts}       
\usepackage{nicefrac}       
\usepackage{microtype}      
\usepackage{lipsum}
\usepackage{graphicx}
\usepackage{dcolumn,longtable,float,comment}
\usepackage{amsmath}
\usepackage{booktabs} 
\usepackage{tabularx} 

\usepackage[backend=biber, style=numeric, sorting=none, url=true, doi=true]{biblatex}
\graphicspath{ {./images/} }

    \makeatletter
\def\@fnsymbol#1{\ensuremath{\ifcase#1\or \dagger\or \dagger\dagger\or
   \mathsection\or \mathparagraph\or \|\or **\or \dagger\dagger
   \or \ddagger\ddagger \else\@ctrerr\fi}}
    \makeatother

\title{An analysis of the effects of open science indicators on citations in the French Open Science Monitor}

\author{
 Giovanni Colavizza\thanks{colavizza@hum.ku.dk} \\
  University of Copenhagen, Denmark\\University of Bologna, Italy\\Odoma LLC, Switzerland\\
  \And
  Lauren Cadwallader \\
  PLOS, United States \\
  \And
  Iain Hrynaszkiewicz\thanks{ihrynaszkiewicz@plos.org} \\
  PLOS, United States \\
}

\begin{document}
\maketitle
\begin{abstract}
This study investigates the correlation of citation impact with various open science indicators (OSI) within the French Open Science Monitor (FOSM), a dataset comprising approximately 900,000 publications authored by French authors from 2020 to 2022. By integrating data from OpenAlex and Crossref, we analyze open science indicators such as the presence of a pre-print, data sharing, and software sharing in 576,537 publications in the FOSM dataset. Our analysis reveals a positive correlation between these OSI and citation counts. Considering our most complete citation prediction model, we find pre-prints are correlated with a significant positive effect of $19\%$ on citation counts, software sharing of $13.5\%$, and data sharing of $14.3\%$. We find large variations in the correlations of OSIs with citations in different research disciplines, and observe that open access status of publications is correlated with a $8.6\%$ increase in citations in our model. While these results remain observational and are limited to the scope of the analysis, they suggest a consistent correlation between citation advantages and open science indicators. Our results may be valuable to policy makers, funding agencies, researchers, publishers, institutions, and other stakeholders who are interested in understanding the academic impacts, or effects, of open science practices.
\end{abstract}

\section{Introduction}
Open science is a broad concept and, as defined by the United Nations Educational Scientific and Cultural Organization (UNESCO), combines various movements and practices aiming to make multilingual scientific knowledge openly available, accessible and reusable for everyone ~\cite{unesco_unesco_2021a}. Within this broad definition of open science is the concept of ``open scientific knowledge'', which includes open research outputs and practices associated, or potentially associated, with scientific publications such as open research data, open code and software, and open access to research publications. These open science practices support increased access to, reuse and reproducibility of research, enable new collaborations, and support greater scrutiny of and trust in research. These assumed, and to some extent, evidenced benefits of open science (also known as open research) are among reasons for various stakeholders -- researchers, governments, policy makers, research funders and institutions, and journals -- to promote open science practices, to various extents. Open science is often promoted through policies and norms of different stakeholder groups, and policies can increase the prevalence of open science practices and outputs -- such as increasing the amount of research publications available open access, and more research data being shared openly. However, even more important to policy makers, practitioners, and beneficiaries of open science is the impact, or effects of these open science practices. Evidence of benefits or consequences of open science and associated policies is needed to guide policy implementation and refinement, and justify further investment in open science activities. Monitoring of open science prevalence, and effects, is therefore of increasing importance to stakeholders, including UNESCO member states, which have adopted its recommendation on open science.

The French Open Science Monitor (FOSM; \url{https://frenchopensciencemonitor.esr.gouv.fr}) is an initiative of the French Ministry of Higher Education and Research and more specifically the French Committee for Open Science, which ensures implementation of the French National Open Science Policy. To enable the French Committee for Open Science to monitor (measure) the impact of its National Open Science Policy, it introduced the FOSM in 2018, to measure the progress in the adoption of open science practices in France. The first edition of the FOSM in 2018 provided indicators on open access to scientific publications. The 2023 edition introduced indicators on sharing of research data, and sharing of code and software~\cite{bassinet_largescale_2023}. The FOSM dataset includes scientific publications -- journal articles, book chapters, doctoral theses -- with at least one French affiliation and a Crossref Digital Object Identifier (DOI), or a Hyper Articles en Ligne (HAL) identifier for publications without a DOI. The inclusion of publications with at least one French affiliation, rather than only publications with a French corresponding affiliation, means the FOSM dataset provides a comprehensive dataset of publications for an entire country, France, approximately 160,000 publications per year. FOSM is based on open data sources and uses open source tools to produce its results. These include (meta)data from Unpaywall, OpenAlex, the Directory of Open Access Journals (DOAJ), PubMed, Crossref, article PDFs and web crawling techniques. The software tools used to produce the FOSM are open source and include GROBID, a tool for extracting structured data (in XML format) from PDFs; DataStet, a tool that identifies mentions of research data and research datasets in publications; and Softcite, which identifies code and software mentions in publications (\url{https://barometredelascienceouverte.esr.gouv.fr/about/methodology}). The methodology of FOSM has been shown to converge for detecting countrywide scientific publications, when compared with alternatives using proprietary data sources.~\cite{10.1162/qss_a_00179} The data produced by the FOSM to create its visualizations and indicators are also made available under an open licence (\url{https://barometredelascienceouverte.esr.gouv.fr/about/opendata}), and institutions, such as the University of Lorraine, have adapted the Monitor to create local versions.

The FOSM -- and initiatives from members of our group such as PLOS Open Science Indicators~\cite{hrynaszkiewicz_plos_2022a} -- is part of a growing number of global initiatives focused on open science monitoring. These initiatives have coalesced  around the global Open Science Monitoring Initiative (OSMI; \url{https://open-science-monitoring.org}), which launched in 2024 and aims to promote the adoption of common principles~\cite{opensciencemonitoringinitiative_principles_2024} and frameworks for open science monitoring globally, along with specifications for their implementation. Monitoring the outcomes (or impacts/consequences) across diverse regions, contexts, and disciplines is equally important to monitoring the prevalence of open science practices.~\cite{rafols_monitoring_2024} Measuring the outcomes and impacts of open science is also generally more challenging at a large scale, as it involves evaluating qualities and long-term effects of policy and behavioral changes in real-world settings. The PathOS (Open Science Impact Pathways; \url{https://pathos-project.eu}) project aimed to improve ways to measure open science impacts, and produced several systematic scoping reviews on the state of the art on the impacts of open science in three areas: academic impacts, societal impacts, and economic impacts.~\cite{cole_societal_2024a,klebel_academic_2025,tsipouri_economic_2025a}

Particularly relevant to our area of study is the systematic scoping review of academic impacts of open science, Klebel \textit{et al}.~\cite{klebel_academic_2025} , which identified citations, quality, efficiency, equity, reuse, ethics and reproducibility as areas of academic impact that have been explored -- along with identifying gaps in the evidence for academic impacts of open science. Although there are limitations to what citations can tell evaluators and users of research, citations remain an important metric for understanding the impact of research findings, and are associated with higher research quality, and researcher recognition. Further, due to their widespread use, citations can be measured across large corpora of articles. Klebel \textit{et al.} \cite{klebel_academic_2025} identified 22 studies reporting an impact of open data (data sharing) on citations; four for pre-prints; and three for open code (code sharing). However, none confirms a causal relationship between open science practices and citation advantages. Of the previous studies on citation advantages for open data, those which explore the effects of sharing research data in a data repository linked to journal articles are most relevant to our research question and methodology. Piwowar and Vision \cite{piwowar_data_2013a} reported a 9\% citation advantage in gene expression studies, Henneken and Accomazzi found a 20\% increase in astronomy studies \cite{henneken_linking_2011a}, and across 531,889 articles published in BMC and PLOS journals, Colavizza \textit{et al.} \cite{colavizza_citation_2020} found a 25\% increase in citations for articles that link to data shared in an external data repository. Fu and Huey reported a 36\% citation advantage for articles that share an associated pre-print \cite{fu_releasing_2019a} and Fraser \textit{et al}., found that after 36 months, articles with an associated pre-print received 1.74 times more citations on average \cite{fraser_relationship_2020a}. Citation effects of code sharing are less well studies across large and/or cross-disciplinary corpora but positive associations of code sharing on citations have been found by others.~\cite{vandewalle_code_2012a,kucharsky_code_2020a} Our previous study published in 2024 complemented and contrasted these results by finding a 20.2\% citation advantage for the sharing of pre-print and 4.3\% advantage for the sharing of data in a repository.~\cite{colavizza_analysis_2024} However, we did not find a statistically significant citation advantage for code sharing. In this previous analysis we used the PLOS Open Science Indicators dataset, which is produced in collaboration with DataSeer, and has monitored five open science practices including data, code, and pre-print sharing, as well as, more recently, protocol sharing, and study registration (preregistration). As an ongoing open science monitoring initiative, the PLOS Open Science Indicators dataset is updated periodically as new PLOS and comparator content is published. Our previous study was based on version 5 of the dataset, which included 121,999 articles, the majority of which are PLOS articles.~\cite{publiclibraryofscience_plos_2023a} 

In this present study, we wanted to explore if similar or different citation effects for the same open science practices we evaluated in our previous study (data sharing, code sharing, pre-print sharing) exist in a larger and more diverse sample of publications. To enable us proceed efficiently we sought openly available data on open science monitoring activities of research publications that could be reused in our citation prediction model. Given the alignment of the indicators measured in the PLOS Open Science Indicators and the indicators included in the FOSM, and its production of reusable open data, we chose the FOSM's larger and more diverse dataset (\url{https://data.enseignementsup-recherche.gouv.fr/explore/dataset/open-access-monitor-france/information}) to explore this question. Furthermore, given the purpose of the FOSM is to inform understanding of open science policy effects and the established national open science policy in France, we assumed the results would be valuable for policy makers, funding agencies, researchers, publishers, institutions, and other stakeholders.

\section{Data and Methods}

The 2023 edition of the FOSM dataset contained approximately 900'000 publications at the time of our analysis. The indicators of interest include whether data and software were created and shared, and whether there are pre-prints associated with published articles. We seek to answer the research question based on the FOSM dataset, namely, whether and to what extent a citation impact premium is received on average by articles following some or all of the open science practices under consideration. This work follows the methodology, and expands upon the results of a previously published work~\cite{colavizza_citation_2020,colavizza_analysis_2024}, including PLOS' Open Science Indicators~\cite{publiclibraryofscience_plos_2023a}. 

To collect the data for analysis, we proceed as follows:
\begin{enumerate}
    \item Acquire the FOSM dataset (\url{https://data.enseignementsup-recherche.gouv.fr/explore/dataset/open-access-monitor-france/information}). We use the December 30, 2022 snapshot (end of coverage period). This snapshot includes 897'426 entries. Of these, 545'158 are classified as journal articles (60.75\%). The publications were published between 2020 and 2022, both years included. The number of publications per year is roughly uniform.
    \item Query OpenAlex (\url{https://openalex.org}). for every DOI present in FOSM, and download the full OpenAlex record locally. OpenAlex was queried between July 12 and 15, 2024. The number of entries in FOSM with an OpenAlex match, after the removal of duplicates, is 576'537 (64.24\%). This dataset contains 479'700 journal articles (83.2\% of the matched total), and is used in what follows.
    \item Add pre-print information from Crossref, using an experimental pre-print-publication relationship dataset created by Crossref.~\cite{tkaczyk_crossref_2023} This dataset covers Crossref publications up to the end of August 2023, therefore fully matching the FOSM coverage. The number of entries in this Crossref dataset is 641'950. Of these, 22'303 match with both FOSM and OpenAlex (3.9\% of FOSM with an OpenAlex match). The total number of pre-print matches, including also arXiv matches indexed in OpenAlex, is 44'763 (7.8\%). This proportion follows known trends in the literature~\cite{levchenko_enabling_2024}.
\end{enumerate}

From OpenAlex, for every entry, we take the month of publication, the number of authors and references, whether there is a pre-print in arXiv, and the citation count at the time of the query (July 2024). From Crossref, we recover possible further matches with other pre-print servers. From FOSM, we take every other variable we use in what follows.

We provide several statistics on the available variables. Please note that the totals may vary since variables differ in the amount of not available values. See Table~\ref{tab:logical_summary} for NA value counts. Table \ref{tab:values_by_category} shows the publication counts for different BSO categories. BSO (Barom\`etre de la Science Ouverte) is a system to classify scientific domains into macro areas. In FOSM, each publication is assigned a BSO class automatically, see \cite{jeangirard_monitoring_2019} for methodological details. Note that Biology (fond.) stands for fundamental (\textit{fondamentale} in French), as distinct from applied biology belonging to another class. Next, we show the publication counts by publication typology (genre) (Table \ref{tab:values_by_type}), showing a large majority for journal articles, by access status (Table \ref{tab:values_by_access_type}) and open access status (Table \ref{tab:values_by_open_access_type}, please note the values here refer only to open access publications). Lastly, we provide descriptive statistics for the main dependent variable (citation counts), controls (Table \ref{tab:summary_statistics}), and open science indicator controls (Table \ref{tab:logical_summary}). Minimum values at zero in Table \ref{tab:summary_statistics} for the number of authors and references of publications likely hint at missing data limitations from OpenAlex. We dealt with such cases in the models that follow by adding one, in order to avoid dropping observations on control, low-importance variables. We note that the maximum correlation among controls is 0.25 for software and data sharing, thus ruling out issues due to multicollinearity. This is part of standard checks done before model fitting, that include the need to avoid using highly correlated controls which would make the model fitting less reliable.

From now on, we consider as open science indicators (OSI) the following variables of interest: pre-print publication, software shared, data shared. We also include software used and created, and data used and created. Although strictly speaking these are not open science practices, they are useful comparisons for sharing activities.

\begin{table}[h!]
\caption{Publication counts by BSO category.}
\label{tab:values_by_category}
\centering
\begin{tabularx}{\textwidth}{>{\raggedright\arraybackslash}Xr}
\toprule
\textbf{BSO category} & \textbf{Publication counts} \\
\midrule
Medical research & 160'672 \\
Biology (fond.) & 95'019 \\
Earth, Ecology, Energy and applied biology & 55'852 \\
Humanities & 51'309 \\
Physical sciences, Astronomy & 48'659 \\
Social sciences & 43'164 \\
Computer and information sciences & 37'948 \\
Engineering & 36'516 \\
Chemistry & 28'856 \\
Mathematics & 18'437 \\
unknown & 105 \\
\bottomrule
\end{tabularx}
\end{table}

\begin{table}[h!]
\caption{Publication counts by typology (genre).}
\label{tab:values_by_type}
\centering
\begin{tabularx}{\textwidth}{>{\raggedright\arraybackslash}Xr}
\toprule
\textbf{Publication genre} & \textbf{Publication counts} \\
\midrule
Journal article & 470'434 \\
Book chapter & 45'795 \\
Proceedings & 29'191 \\
Other & 18'127 \\
Pre-print & 9'636 \\
Book & 3'354 \\
\bottomrule
\end{tabularx}
\end{table}

\begin{table}[h!]
\caption{Publication counts by access status.}
\label{tab:values_by_access_type}
\centering
\begin{tabularx}{\textwidth}{>{\raggedright\arraybackslash}Xr}
\toprule
\textbf{Access status} & \textbf{Publication counts} \\
\midrule
Closed & 193'222 \\
Publisher-repository & 183'554 \\
Repository & 102'907 \\
Publisher & 96'534 \\
\bottomrule
\end{tabularx}
\end{table}

\begin{table}[h!]
\caption{Publication counts by open access status.}
\label{tab:values_by_open_access_type}
\centering
\begin{tabularx}{\textwidth}{>{\raggedright\arraybackslash}Xr}
\toprule
\textbf{Open access status} & \textbf{Publication counts} \\
\midrule
Gold & 132'435 \\
Green & 107'370 \\
Hybrid & 92'824 \\
Bronze & 51'329 \\
\bottomrule
\end{tabularx}
\end{table}

\begin{table}[h!]
\caption{Descriptive statistics for the dependent variable and a set of publication and author level controls.}
\label{tab:summary_statistics}
\centering
\begin{tabularx}{\textwidth}{>{\raggedright\arraybackslash}Xrrr}
\toprule
\textbf{Statistic} & \textbf{Cited by Count} & \textbf{Number of Authors} & \textbf{Number of References} \\
\midrule
Min.    & 0   & 0   & 0    \\
1st Qu. & 0   & 2   & 1    \\
Median  & 2   & 5   & 23   \\
Mean    & 11  & 6.9   & 32   \\
3rd Qu. & 10  & 8   & 46   \\
Max.    & 63'705 & 100 & 4'083  \\
\bottomrule
\end{tabularx}
\end{table}

\begin{table}[h!]
\caption{Descriptive statistics for open science indicators.}
\label{tab:logical_summary}
\centering
\begin{tabularx}{\textwidth}{>{\raggedright\arraybackslash}Xrrr}
\toprule
\textbf{Variable} & \textbf{FALSE Count} & \textbf{TRUE Count} & \textbf{NA Count} \\
\midrule
Pre-print match & 531'774 & 44'763 & 0 \\
Software used & 202'336 & 136'300 & 237'901 \\
Software created & 318'136 & 20'500 & 237'901 \\
Software shared & 330'305 & 8'331 & 237'901 \\
Data used & 81'890 & 256'201 & 238'446 \\
Data created & 234'950 & 103'141 & 238'446 \\
Data shared & 303'080 & 35'011 & 238'446 \\
Is OA & 193'222 & 382'995 & 320 \\
\bottomrule
\end{tabularx}
\end{table}

\begin{figure}[H]
\caption{Adoption of open science indicators (OSI) over time in FOSM. Each OSI remains adopted by a fraction of publications, with a stable outlook in the recent years covered by the dataset.}\label{fig:OSI_time}
\centering
\includegraphics[width=0.9\textwidth]{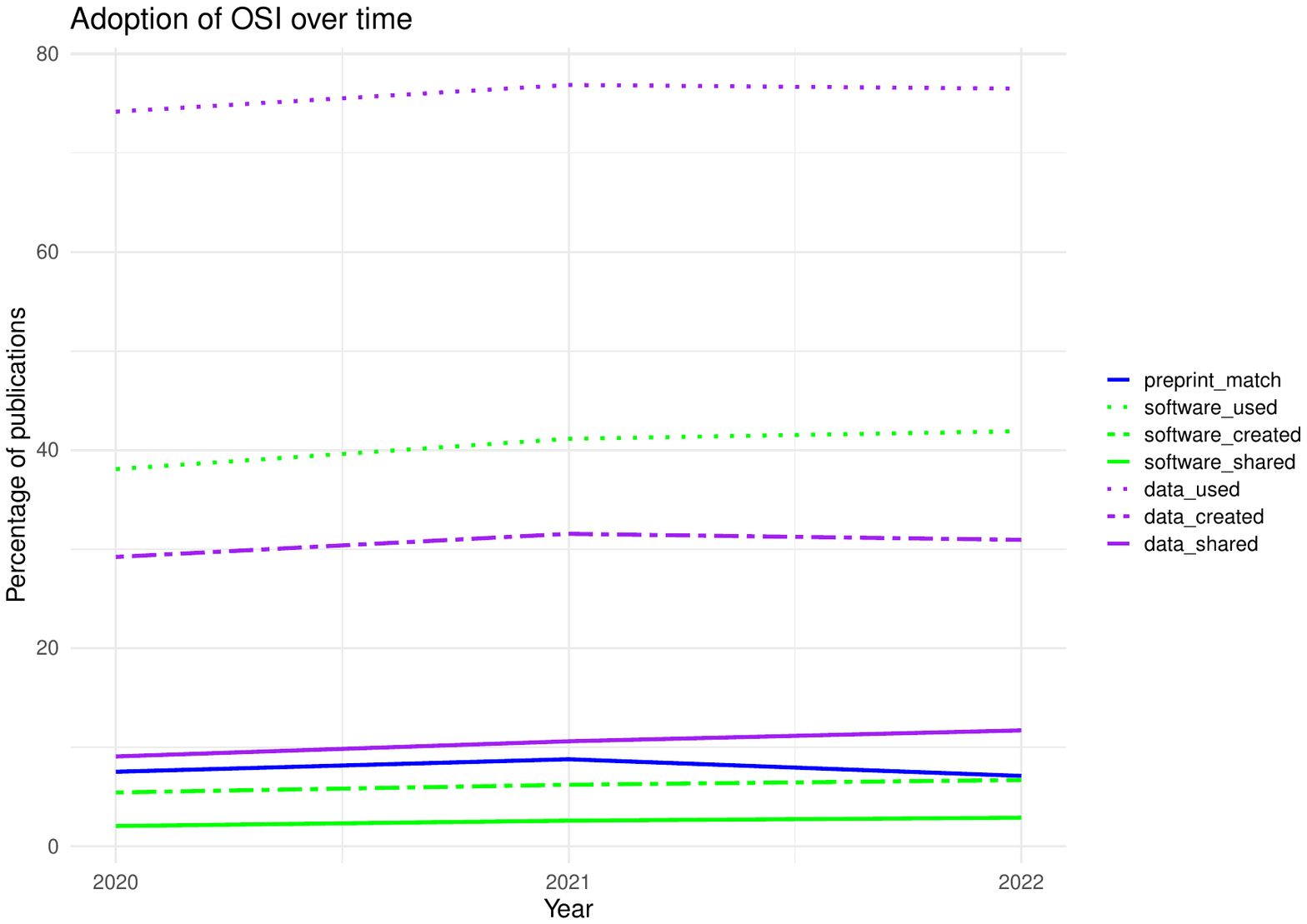}
\end{figure}

\section{Results}

\begin{figure}[H]
\caption{Adoption of OSI by BSO class, as shown in Table~\ref{tab:values_by_category}. Each OSI remains adopted by a fraction of publications, and there is significant variation across domains.}\label{fig:OSI_division}
\centering
\includegraphics[width=0.9\textwidth]{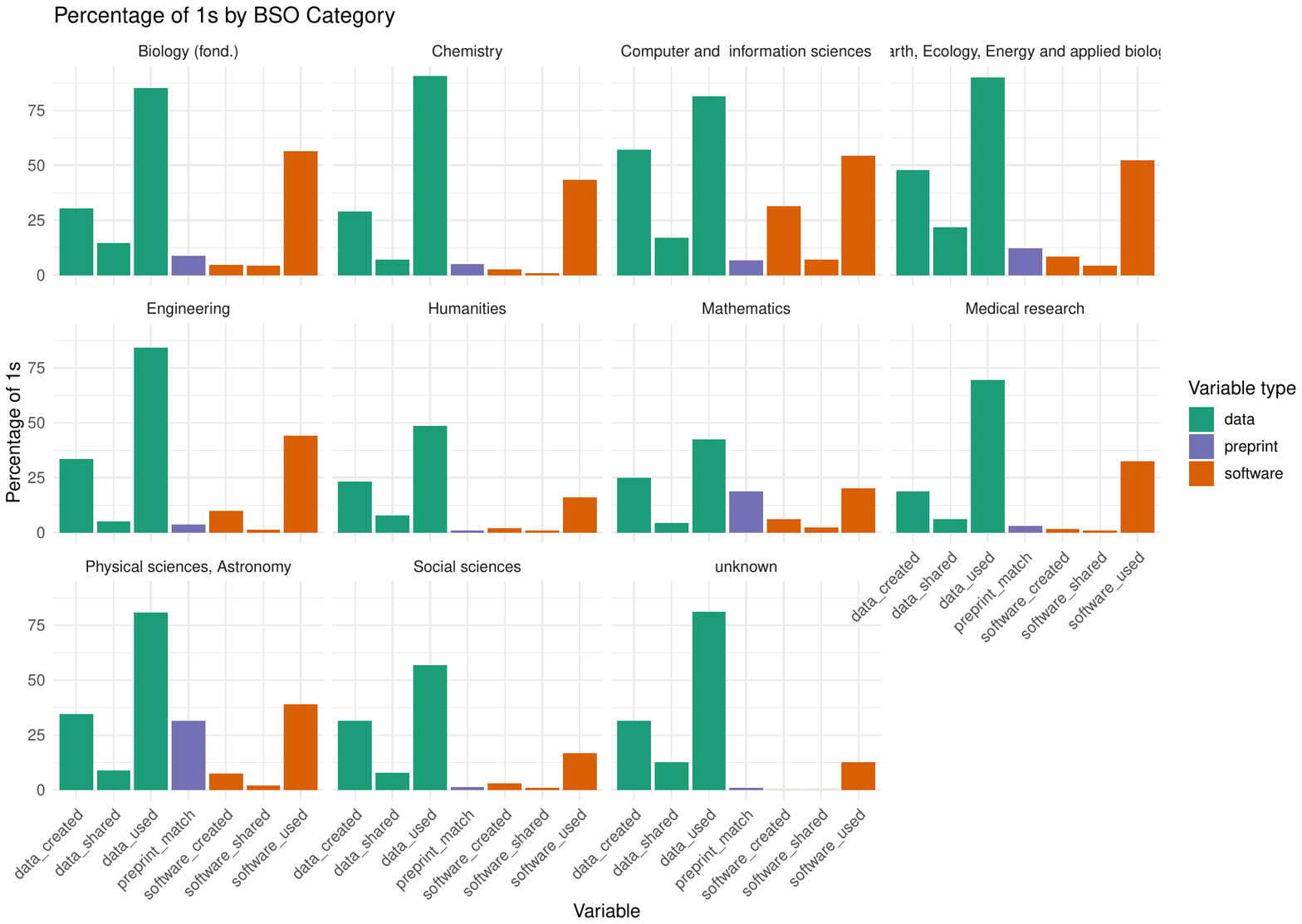}
\end{figure}

We start by visualizing the temporal trends in OSI adoption, reminding the reader that the FOSM dataset used in our analysis  only covers publication from 2020, 2021, and 2022. In Figure~\ref{fig:OSI_time}, we notice how most OSI remain stable and relatively low over time, despite significant levels of data and software use. In Figure~\ref{fig:OSI_division}, we unpack trends by BSO class, noticing some common trends such as higher software creation in Computer Science, and pre-print adoption in Physics and Mathematics.

\subsection{Modeling}

The base model we use is described in Equation~\ref{eq:without_division}, and the full model is described in Equation~\ref{eq:with_division}. Variable transformations are shown, numerical variables are given in Italics, and categorical variables are in regular text. Variables are grouped along lines. This model follows closely from our previous work, where an illustration of the assumed causal dependency graph among variable groups is also provided~\cite{colavizza_analysis_2024}. We remark that, while the variables year and month could also be considered as categorical or fixed effects, the results do not differ substantially. The code to replicate these and all other models is provided. We also notice that, with respect to our previous work, we have no ability to control for the mean h-index of the authors at the time of publication, and we use the BSO classification instead of the Fields of Research classification codes (\url{https://www.arc.gov.au/manage-your-grant/classification-codes-rfcd-seo-and-anzsic-codes}). At the same time, we add further variables such as software/data used or created, publication genre (e.g. Journal article, Book chapter), and open access status. The absence of the h-index control is the most significant limitation of our work here, in terms of its comparability to our previous work, as it constitutes a proxy for author success or reputation at the time of publication, which might also correlate with resource availability. To confirm the absence of discrepancies, we provide results on a small set of journal articles that are part of both PLOS OSI and FOSM in the Appendix. 

The codebase contains a variety of post-estimation diagnostics and alternative models, including residual behavior, influential observations, multicollinearity, model specification checks. A Breusch–Pagan test strongly rejects homoskedasticity (BP = $9888.8$, p < $0.001$). We therefore also add a robust OLS model throughout. The codebase also contains ways to model with robust and clustered standard errors as further checks. These are omitted here as they align with the provided results.

\begin{equation}
\begin{aligned}
\log(n\_cit\_tot + 1) = & \log(n\_authors + 1) + \log(n\_references + 1) + year + month + \\
             & \text{pre-print\_match} + \\
             & \text{software\_created} + \text{software\_shared} + \\
             & \text{data\_created} + \text{data\_shared}
\end{aligned}
\label{eq:without_division}
\end{equation}

\begin{equation}
\begin{aligned}
\log(n\_cit\_tot + 1) = & \log(n\_authors + 1) + \log(n\_references + 1) + year + month + \\
             & \text{genre} + \text{pre-print\_match} + \\
             & \text{software\_used} + \text{software\_created} + \text{software\_shared} + \\
             & \text{data\_used} + \text{data\_created} + \text{data\_shared} + \\
             & \text{is\_oa} + \text{bso\_classification}
\end{aligned}
\label{eq:with_division}
\end{equation}

Starting with the base model in Table~\ref{tab:model_base}, we provide results for an Ordinary Least Squares (OLS) model and a robust linear model as a comparison. Boolean variables use ``False'' as reference category. The results are aligned and show a relatively high explained variance with the base model having $R^2 = .45$. The model shows expected trends with respect to common controls (authors, references, year of publication), and interesting novel trends with respect to OSI. The OSI show that there is a significant and positive correlation of pre-prints ($22.6\%$), sharing data ($14.1\%$) and sharing software ($6.5\%$) with increased citations. These percentage changes for log-linear relationships are calculated as follows: $(\exp(.204) - 1) \times 100 \approx 22.6\%$. Our next question is whether these results hold when we account for large disciplinary variations in the adoption of open science practices, as well as further controls which we add next.

\begin{longtable}{@{\extracolsep{5pt}}lD{.}{.}{-3} D{.}{.}{-3} } 

  \caption{Results (base model)} 
  \label{tab:model_base}  
 
\\[-1.8ex]\hline 
\hline \\[-1.8ex] 
 & \multicolumn{2}{c}{\textit{Dependent variable:}} \\ 
\cline{2-3} 
\\[-1.8ex] & \multicolumn{2}{c}{n\_cit\_tot\_log} \\ 
\\[-1.8ex] & \multicolumn{1}{c}{\textit{OLS}} & \multicolumn{1}{c}{\textit{robust}} \\ 
 & \multicolumn{1}{c}{\textit{}} & \multicolumn{1}{c}{\textit{linear}} \\ 
\\[-1.8ex] & \multicolumn{1}{c}{(1)} & \multicolumn{1}{c}{(2)}\\ 
\hline \\[-1.8ex] 
 n\_authors\_log & 0.399^{***} & 0.386^{***} \\ 
  & (0.002) & (0.002) \\ 
  & & \\ 
 n\_references\_log & 0.454^{***} & 0.468^{***} \\ 
  & (0.001) & (0.001) \\ 
  & & \\ 
 year & -0.330^{***} & -0.324^{***} \\ 
  & (0.002) & (0.002) \\ 
  & & \\ 
 month & -0.023^{***} & -0.023^{***} \\ 
  & (0.0005) & (0.0004) \\ 
  & & \\ 
 C(pre-print\_match) & 0.204^{***} & 0.195^{***} \\ 
  & (0.005) & (0.005) \\ 
  & & \\ 
 C(software\_created) & 0.028^{***} & 0.029^{***} \\ 
  & (0.007) & (0.007) \\ 
  & & \\ 
 C(software\_shared) & 0.063^{***} & 0.063^{***} \\ 
  & (0.011) & (0.011) \\ 
  & & \\ 
 C(data\_created) & 0.127^{***} & 0.122^{***} \\ 
  & (0.004) & (0.004) \\ 
  & & \\ 
 C(data\_shared) & 0.132^{***} & 0.127^{***} \\ 
  & (0.006) & (0.006) \\ 
  & & \\ 
 Constant & 666.250^{***} & 653.919^{***} \\ 
  & (4.173) & (4.081) \\ 
  & & \\ 
\hline \\[-1.8ex] 
Observations & \multicolumn{1}{c}{337,929} & \multicolumn{1}{c}{337,929} \\ 
R$^{2}$ & \multicolumn{1}{c}{0.450} &  \\ 
Adjusted R$^{2}$ & \multicolumn{1}{c}{0.450} &  \\ 
Residual Std. Error (df = 337919) & \multicolumn{1}{c}{0.970} & \multicolumn{1}{c}{0.880} \\ 
F Statistic & \multicolumn{1}{c}{30,682.610$^{***}$ (df = 9; 337919)} &  \\ 
\hline 
\hline \\[-1.8ex] 
\textit{Note:}  & \multicolumn{2}{r}{$^{*}$p$<$0.1; $^{**}$p$<$0.05; $^{***}$p$<$0.01} \\ 
\end{longtable} 

In the full model we propose, in Table~\ref{tab:model_full}, we add the following variables: publication genre, software/data used, open access status, language, and BSO classification. We take the categories with most publications as reference category, specifically: `journal-article' for genre, `English' for language, and `Medical research' for BSO classification. Our model confirms previous trends, namely pre-print match contributing to a significant and positive correlation of $19\%$ with citation counts, software sharing of $13.5\%$, and data sharing of $14.3\%$. We also find positive effects from software/data creation. Publications where data is used tend to be more cited, whilst those when software is used slightly less so. Publications in English are significantly more cited than publications in other languages. Disciplinary differences (with respect to the baseline Medical research), and publication genre differences (with respect to the baseline journal article), also follow known trends.

While open access (OA) status is not a relevant indicator for the PLOS OSI dataset given all content is OA, OA status is an important aspect of the FOSM, with OA adoption reaching 65.6\% in 2022 in French publications (https://frenchopensciencemonitor.esr.gouv.fr/). We included OA status in our model and can observe a $8.6\%$ citation increase correlated with publications that are OA compared to those that are closed access. For a review on the citation impact of open access status, see \cite{langham-putrow_open_2021a}. OA status can be explored in more depth considering OA typologies using our data and code.~\cite{colavizza_study_2025}

\begin{longtable}{@{\extracolsep{5pt}}lD{.}{.}{-3} D{.}{.}{-3} } 

  \caption{Results (full model)} 
  \label{tab:model_full} 

\\[-1.8ex]\hline 
\hline \\[-1.8ex] 
 & \multicolumn{2}{c}{\textit{Dependent variable:}} \\ 
\cline{2-3} 
\\[-1.8ex] & \multicolumn{2}{c}{n\_cit\_tot\_log} \\ 
\\[-1.8ex] & \multicolumn{1}{c}{\textit{OLS}} & \multicolumn{1}{c}{\textit{robust}} \\ 
 & \multicolumn{1}{c}{\textit{}} & \multicolumn{1}{c}{\textit{linear}} \\ 
\\[-1.8ex] & \multicolumn{1}{c}{(1)} & \multicolumn{1}{c}{(2)}\\ 
\hline \\[-1.8ex] 
 n\_authors\_log & 0.346^{***} & 0.326^{***} \\ 
  & (0.003) & (0.002) \\ 
  & & \\ 
 n\_references\_log & 0.379^{***} & 0.382^{***} \\ 
  & (0.001) & (0.001) \\ 
  & & \\ 
 year & -0.327^{***} & -0.316^{***} \\ 
  & (0.002) & (0.002) \\ 
  & & \\ 
 month & -0.023^{***} & -0.022^{***} \\ 
  & (0.0004) & (0.0004) \\ 
  & & \\ 
 C(genre)book & 0.635^{***} & 0.604^{***} \\ 
  & (0.032) & (0.032) \\ 
  & & \\ 
 C(genre)book-chapter & -0.686^{***} & -0.668^{***} \\ 
  & (0.010) & (0.010) \\ 
  & & \\ 
 C(genre)other & -0.541^{***} & -0.502^{***} \\ 
  & (0.013) & (0.013) \\ 
  & & \\ 
 C(genre)preprint & -1.392^{***} & -1.423^{***} \\ 
  & (0.011) & (0.011) \\ 
  & & \\ 
 C(genre)proceedings & -0.534^{***} & -0.541^{***} \\ 
  & (0.010) & (0.010) \\ 
  & & \\ 
 C(preprint\_match) & 0.174^{***} & 0.177^{***} \\ 
  & (0.006) & (0.005) \\ 
  & & \\ 
 C(software\_used) & -0.057^{***} & -0.030^{***} \\ 
  & (0.004) & (0.004) \\ 
  & & \\ 
 C(software\_created) & 0.061^{***} & 0.053^{***} \\ 
  & (0.007) & (0.007) \\ 
  & & \\ 
 C(software\_shared) & 0.127^{***} & 0.116^{***} \\ 
  & (0.011) & (0.011) \\ 
  & & \\ 
 C(data\_used) & 0.148^{***} & 0.162^{***} \\ 
  & (0.005) & (0.005) \\ 
  & & \\ 
 C(data\_created) & 0.076^{***} & 0.075^{***} \\ 
  & (0.004) & (0.004) \\ 
  & & \\ 
 C(data\_shared) & 0.134^{***} & 0.122^{***} \\ 
  & (0.006) & (0.006) \\ 
  & & \\ 
 C(is\_oa) & 0.083^{***} & 0.075^{***} \\ 
  & (0.005) & (0.005) \\ 
  & & \\ 
 C(lang\_reduce)fr & -0.584^{***} & -0.538^{***} \\ 
  & (0.007) & (0.006) \\ 
  & & \\ 
 C(lang\_reduce)other & -0.562^{***} & -0.515^{***} \\ 
  & (0.014) & (0.014) \\ 
  & & \\ 
 C(bso\_classification)Biology (fond.) & 0.048^{***} & 0.063^{***} \\ 
  & (0.005) & (0.005) \\ 
  & & \\ 
 C(bso\_classification)Chemistry & 0.061^{***} & 0.095^{***} \\ 
  & (0.007) & (0.007) \\ 
  & & \\ 
 C(bso\_classification)Computer and  information sciences & 0.107^{***} & 0.118^{***} \\ 
  & (0.008) & (0.008) \\ 
  & & \\ 
 C(bso\_classification)Earth, Ecology, Energy and applied biology & 0.026^{***} & 0.051^{***} \\ 
  & (0.006) & (0.006) \\ 
  & & \\ 
 C(bso\_classification)Engineering & 0.066^{***} & 0.099^{***} \\ 
  & (0.007) & (0.007) \\ 
  & & \\ 
 C(bso\_classification)Humanities & 0.033^{***} & 0.051^{***} \\ 
  & (0.008) & (0.008) \\ 
  & & \\ 
 C(bso\_classification)Mathematics & -0.140^{***} & -0.115^{***} \\ 
  & (0.009) & (0.009) \\ 
  & & \\ 
 C(bso\_classification)Physical sciences, Astronomy & -0.102^{***} & -0.059^{***} \\ 
  & (0.007) & (0.006) \\ 
  & & \\ 
 C(bso\_classification)Social sciences & 0.095^{***} & 0.099^{***} \\ 
  & (0.008) & (0.008) \\ 
  & & \\ 
 C(bso\_classification)unknown & -0.259 & -0.221 \\ 
  & (0.231) & (0.226) \\ 
  & & \\ 
 Constant & 660.758^{***} & 638.442^{***} \\ 
  & (3.973) & (3.888) \\ 
  & & \\ 
\hline \\[-1.8ex] 
Observations & \multicolumn{1}{c}{337,928} & \multicolumn{1}{c}{337,928} \\ 
R$^{2}$ & \multicolumn{1}{c}{0.503} &  \\ 
Adjusted R$^{2}$ & \multicolumn{1}{c}{0.503} &  \\ 
Residual Std. Error (df = 337898) & \multicolumn{1}{c}{0.922} & \multicolumn{1}{c}{0.858} \\ 
F Statistic & \multicolumn{1}{c}{11,779.980$^{***}$ (df = 29; 337898)} &  \\ 
\hline 
\hline \\[-1.8ex] 
\textit{Note:}  & \multicolumn{2}{r}{$^{*}$p$<$0.1; $^{**}$p$<$0.05; $^{***}$p$<$0.01} \\ 
\end{longtable}

These results are robust to a variety of modeling changes, which we include in our accompanying repository.

\begin{figure}[H]
\caption{Percentage change on citation counts linked to each OSI, by BSO class. We only consider journal articles to fit these models.}\label{fig:BSO_table}
\centering
\includegraphics[width=0.99\textwidth]{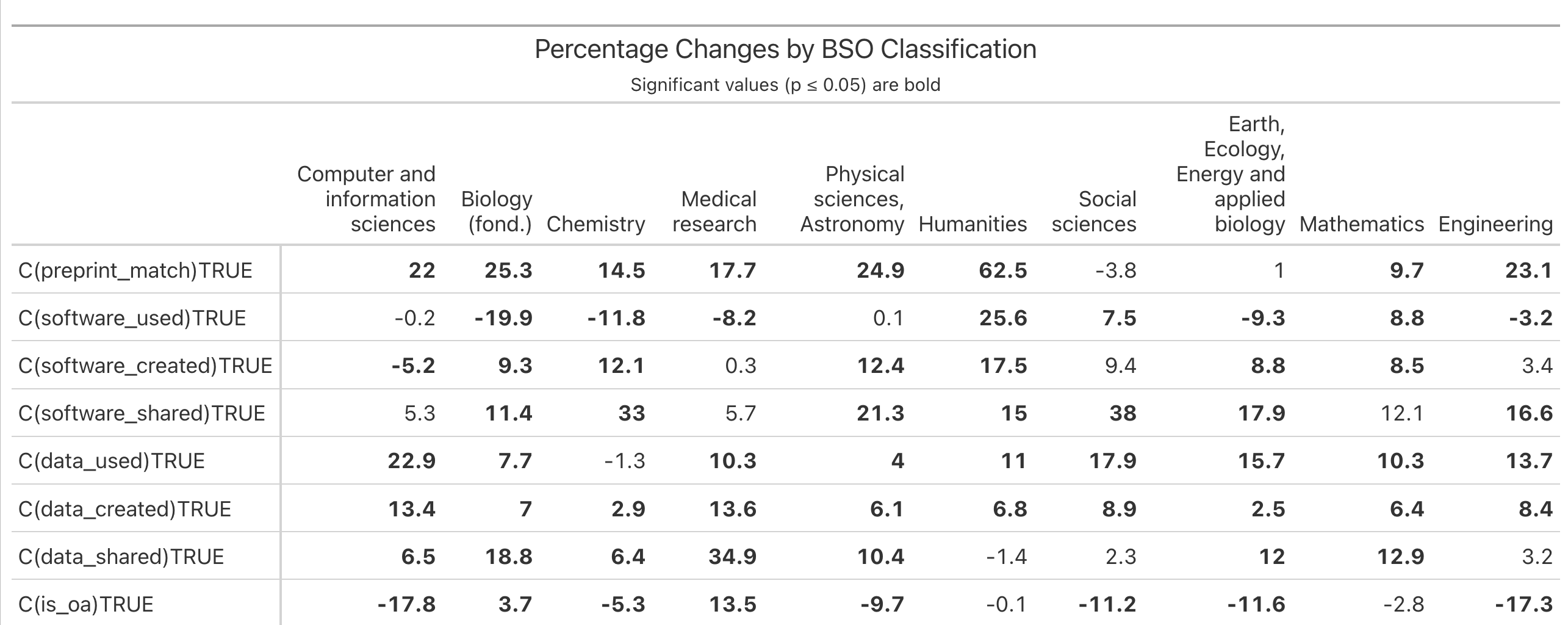}
\end{figure}

Lastly, we provide a summary table with the percentage change on citation counts linked to each OSI available, organizing publications by scientific domain (BSO class) and using only the main publication genre: journal articles. Percentage changes are calculated exactly as we did above, from discipline-specific log-linear models, with the reference category being publications within the same BSO class not exhibiting the given open science indicator. These results should be more easily readable than those provided above, when interested in differences across scientific domains. We fit a full model specification for every BSO class, and provide results in Figure~\ref{fig:BSO_table}. While these results confirm the trends discussed above, they also highlight significant variations across scientific domains. For example, the open access status is positively correlated to citation counts only in Medicine and Biology, and negatively elsewhere. 

\section{Discussion and Conclusions}

In this work, we have explored the FOSM dataset and investigated whether three open science indicators (OSI) -- data sharing, code sharing, pre-print posting -- are correlated with a citation advantage received by the publications that exhibit these open science practices. To our knowledge, this is the largest scale analysis of these open science practices and citations to date, and the first such study offering a nationwide (French) perspective on this topic. We use over half a million publications from all scientific domains, and find a positive correlation of these open science practices with citation counts. This study is observational and limited by the scope of the analysis and the available data, therefore its results should be taken with caution. As a reanalysis of existing data, which have been generated with machine learning approaches, our results are also limited by the accuracy of the original data, in particular its detection of shared data and code for articles. The accuracy (F1 scores) of the machine learning methodology of the FOSM for detection of data and code use, creation, and sharing range from 82.19 to 99.03 are fully reported in [2]. We are in no position to establish causality nor to control for additional, confounding effects. Significant confounding effects that we could not consider are the quality of the research, and the reputation of the authors (as measured by h-index) and of the venue of publication. A more developed discussion of these and other limitations of this approach to measuring citation effects is provided in \cite{colavizza_analysis_2024}. In theory, the observed correlation between increased citations and these open science practices in our analysis are cumulative, which would suggest a substantial average increase -- $46.8\%$ -- in citations where an article exhibited all three open science practices. However, such an assumption should be made with caution given the aforementioned limitations.

Our finding of a significant positive effect of pre-print posting on citations, of  $19\%$ in this study, confirms results of previous studies.~\cite{10.7554/eLife.52646,colavizza_analysis_2024} Our finding of a significant positive effect of software sharing of $13.5\%$ contrasts the result of our previous study, which found no statistically significant increase in citations correlated with code (software) sharing.~\cite{colavizza_analysis_2024} This may be due to the larger and more diverse, with respect to research topics and disciplines, dataset used in this study compared to our previous work. However, other studies~\cite{vandewalle_code_2012a} have found as much as a three-fold increase in citations associated with code sharing. Our finding of a significant positive effect of data sharing of $14.3\%$ confirms a significant positive effect observed in two previous studies, which found increases in citations correlated with data sharing in a repository of $25.3\%$ when analysing publications in PLOS and BMC (BioMed Central) and $4.3\%$ when analyzing publications in PLOS OSI.~\cite{colavizza_citation_2020,colavizza_analysis_2024} The larger effect of OSI on citations observed in the current study may be due to the larger and more diverse corpus of publications, which includes, for example, a higher proportion of content from the Humanities and Physical Sciences than these previous two studies. Also, the FOSM publication corpus includes content that is not open access, in contrast to these previous two studies. The source of publications and citation counts (OpenAlex) in the current study is different to our previous studies, which relied on PubMed Central. OpenAlex is emerging as a credible alternative to established databases for bibliometric analyses but may lag behind in the number of indexed references (and as a consequence, citations) -- at least when compared to the commercial database, Scopus.~\cite{alperin_analysis_2024} 

While we have focused our interpretation on average citation effects across the analysed dataset, we can observe some notable differences in these averages when results are segmented by BSO class (research topic or discipline), in Figure~\ref{fig:BSO_table}. For example, data sharing is correlated with a $14.3\%$ increase in citations on average in our analysis but with publications classified as Medical research, we observe a larger,  $34.9\%$ increase. Also, the effect of pre-printing is particularly high in the Humanities, where pre-printing is correlated with a $62.5\%$ increase, compared to $19\%$ in the dataset overall. We can only speculate here on the reasons for these differences. Public data sharing in Medical research can be more challenging compared to fields that do not produce or analyze sensitive data, and so sharing these data comes with additional processing, security, and management costs. Where shared datasets are available, such as from important clinical trials, then it is possible they will have greater reuse value, in turn driving more citations. In other words, we could speculate that the additional effort on behalf of researchers to share high value datasets is rewarded with greater impact of their work. In the Humanities, our sample of publications is smaller than other disciplines covered in the dataset, but we can speculate that cultural and temporal differences in publishing and citation in the Humanities relative to other disciplines are a factor. That is, publication output and speed tends to be slower than the life and medical sciences, and total numbers of citations lower. Therefore, the small proportion of publications that do have a pre-print available could potentially benefit from greater visibility and therefore citations, amplifying the observed average effect of pre-printing relative to disciplinary norms. Conversely, we observe a small negative correlation of data sharing  (-$1.4\%$) in the Humanities.

Regarding the larger difference we observed in the citation effect of data sharing compared to our previous study finding a $4.3\%$ positive correlation, differences in how open science practices are defined and measured between open science monitoring methods could be a factor. In~\cite{colavizza_analysis_2024}, the positive correlation was observed for a specific method of data sharing, data sharing in a repository linked to the article (as opposed to sharing data in online supplementary material). There are some subtle but important differences in how open science practices are defined and, consequently, measured, in different open science monitoring initiatives and software tools. Systematically mapping OSI definitions and measures across multiple studies is outside of the scope of the current work, although we provide a small-scale mapping of a subset of publications between our current and previous study in the Appendix. This issue highlights the need for development of common frameworks for, and consensus on, methods for open science monitoring, using open methods and data. 

With the above limitations, this study adds to the evidence that OSI (open science practices) are associated with increased academic impact of research through citations, and provide for the first time, to our knowledge, a perspective on this question for an entire country, France. Open science monitoring is still in its relative infancy, but France provides a valuable case study having implemented a comprehensive open science policy for research publications and outputs since 2018 and latterly introducing a means to monitor its progress. The FOSM has helped to demonstrate that the prevalence of open science practices such as open access to publications, open data, and open code/software have increased since the implementation of their national policy. Open access to publications has, for example, increased from $38\%$ in 2018 to $64\%$ in 2023 and open data from and $16\%$ to $25\%$ in the same period of time (\url{https://frenchopensciencemonitor.esr.gouv.fr}). Our findings complement these observations by providing policy makers and other stakeholders with evidence that, as well as increasing prevalence of open science, there open science policies may be a benefit in terms of greater impact of research through open science practices as measured by citations. Future research could seek to understand if open science practices lead to other (non-citation based) academic impacts of research, such as whether they lead to increased (re)use of research outputs, improved reproducibility, economic impact, or greater trust in research outputs. More work on determining causal mechanisms of changes in academic impacts of open science is also needed.

\section*{Data and Code Availability}

Please find all data and code to reproduce our results at \url{https://doi.org/10.6084/m9.figshare.27822663}.~\cite{colavizza_study_2025}

\section*{Acknowledgments}
For their support in using the French Open Science Monitor dataset, we thank Eric Jeangirard and Anne L’Hote at the French Ministry of Higher Education and Research, Laetitia Bracco at the University of Lorraine and Laurent Romary at the National Institute for Research in Digital Science and Technology (Inria). We also thank Marin Dacos and Arianna Caporali at the French Ministry of Higher Education and Research for participating in discussions with PLOS that contributed to the conceptualization of this research.

\section*{Author contributions}
GC: Conceptualization, Data curation, Formal analysis, Investigation, Methodology, Software, Supervision, Writing – original draft, Writing – review \& editing. \\
LC: Formal Analysis, Visualization, Writing – review \& editing. \\
IH: Conceptualization, Funding acquisition, Methodology, Project administration, Resources, Supervision, Writing – original draft, Writing – review \& editing.

\section*{Competing interests}
Two of the authors (LC and IH) were at the time of publication employed by PLOS, which produces PLOS Open Science Indicators. PLOS (IH) is a founding member and initiator of the Open Science Monitoring Initiative (OSMI). GC declares no competing interests.

\section*{Funding information}
PLOS provided funding for data acquisition, modeling, and analysis, and had a role in the design, analysis, and preparation of the study. PLOS also provided support in the form of salaries for authors LC and IH.
 
\section*{Appendix}

We report here on an analysis using the same models but focused on the overlap between FOSM and the PLOS OSI dataset (version 7). The overlap consists of 3120 publications, from PLOS and the control group. Note that some control variables (genre, open access status) had to be omitted due to missing factors. The results align with the findings of \cite{colavizza_analysis_2024} and with the findings of this analysis, showing larger effects for data and software sharing. 

Please note there are some differences in definitions and measures in the FOSM compared to PLOS OSI, reflecting the different methods used by the FOSM, and PLOS/DataSeer teams. For reference, the FOSM uses the following definitions for data/software use, creation, and sharing~\cite{bassinet_largescale_2023}:
\begin{itemize}
    \item Data/software use refers to publications that mention the usage of at least one dataset/software.
    \item Data/software created refers to publications that mention the usage and the production of at least one of their datasets/software.
    \item Data/software shared refers to publications that mention the usage, the production and the sharing of at least one of their datasets/software.
\end{itemize}
While our measures of code sharing are well aligned across FOSM and PLOS OSI, in our previous study of PLOS OSI datasets, for data sharing, we assessed the effects of sharing research data on citations in a more specific way, that is, sharing via an online data repository~\cite{colavizza_analysis_2024}.

\begin{longtable}{@{\extracolsep{5pt}}lD{.}{.}{-3} D{.}{.}{-3} } 
  \caption{Base model, PLOS OSI and FOSM overlap.} 
  \label{} 

\\[-1.8ex]\hline 
\hline \\[-1.8ex] 
 & \multicolumn{2}{c}{\textit{Dependent variable:}} \\ 
\cline{2-3} 
\\[-1.8ex] & \multicolumn{2}{c}{n\_cit\_tot\_log} \\ 
\\[-1.8ex] & \multicolumn{1}{c}{\textit{OLS}} & \multicolumn{1}{c}{\textit{robust}} \\ 
 & \multicolumn{1}{c}{\textit{}} & \multicolumn{1}{c}{\textit{linear}} \\ 
\\[-1.8ex] & \multicolumn{1}{c}{(1)} & \multicolumn{1}{c}{(2)}\\ 
\hline \\[-1.8ex] 
 n\_authors\_log & 0.433^{***} & 0.399^{***} \\ 
  & (0.027) & (0.027) \\ 
  & & \\ 
 n\_references\_log & 0.333^{***} & 0.335^{***} \\ 
  & (0.031) & (0.031) \\ 
  & & \\ 
 year & -0.451^{***} & -0.460^{***} \\ 
  & (0.020) & (0.019) \\ 
  & & \\ 
 month & -0.039^{***} & -0.040^{***} \\ 
  & (0.005) & (0.004) \\ 
  & & \\ 
 C(pre-print\_match) & 0.254^{***} & 0.253^{***} \\ 
  & (0.036) & (0.035) \\ 
  & & \\ 
 C(software\_created) & -0.069 & -0.043 \\ 
  & (0.064) & (0.062) \\ 
  & & \\ 
 C(software\_shared) & 0.183^{***} & 0.163^{**} \\ 
  & (0.068) & (0.066) \\ 
  & & \\ 
 C(data\_created) & 0.066^{**} & 0.055^{*} \\ 
  & (0.033) & (0.032) \\ 
  & & \\ 
 C(data\_shared) & 0.158^{***} & 0.158^{***} \\ 
  & (0.047) & (0.045) \\ 
  & & \\ 
 Constant & 912.210^{***} & 928.924^{***} \\ 
  & (39.459) & (38.540) \\ 
  & & \\ 
\hline \\[-1.8ex] 
Observations & \multicolumn{1}{c}{2,812} & \multicolumn{1}{c}{2,812} \\ 
R$^{2}$ & \multicolumn{1}{c}{0.269} &  \\ 
Adjusted R$^{2}$ & \multicolumn{1}{c}{0.267} &  \\ 
Residual Std. Error (df = 2802) & \multicolumn{1}{c}{0.832} & \multicolumn{1}{c}{0.747} \\ 
F Statistic & \multicolumn{1}{c}{114.584$^{***}$ (df = 9; 2802)} &  \\ 
\hline 
\hline \\[-1.8ex] 
\textit{Note:}  & \multicolumn{2}{r}{$^{*}$p$<$0.1; $^{**}$p$<$0.05; $^{***}$p$<$0.01} \\ 
\end{longtable}

\begin{longtable}{@{\extracolsep{5pt}}lD{.}{.}{-3} D{.}{.}{-3} } 
  \caption{Full model, PLOS OSI and FOSM overlap.} 
  \label{} 

\\[-1.8ex]\hline 
\hline \\[-1.8ex] 
 & \multicolumn{2}{c}{\textit{Dependent variable:}} \\ 
\cline{2-3} 
\\[-1.8ex] & \multicolumn{2}{c}{n\_cit\_tot\_log} \\ 
\\[-1.8ex] & \multicolumn{1}{c}{\textit{OLS}} & \multicolumn{1}{c}{\textit{robust}} \\ 
 & \multicolumn{1}{c}{\textit{}} & \multicolumn{1}{c}{\textit{linear}} \\ 
\\[-1.8ex] & \multicolumn{1}{c}{(1)} & \multicolumn{1}{c}{(2)}\\ 
\hline \\[-1.8ex] 
 n\_authors\_log & 0.442^{***} & 0.407^{***} \\ 
  & (0.028) & (0.028) \\ 
  & & \\ 
 n\_references\_log & 0.323^{***} & 0.328^{***} \\ 
  & (0.033) & (0.032) \\ 
  & & \\ 
 year & -0.451^{***} & -0.460^{***} \\ 
  & (0.020) & (0.019) \\ 
  & & \\ 
 month & -0.038^{***} & -0.040^{***} \\ 
  & (0.005) & (0.005) \\ 
  & & \\ 
 C(pre-print\_match) & 0.246^{***} & 0.242^{***} \\ 
  & (0.037) & (0.036) \\ 
  & & \\ 
 C(software\_used) & 0.022 & 0.016 \\ 
  & (0.040) & (0.039) \\ 
  & & \\ 
 C(software\_created) & -0.084 & -0.055 \\ 
  & (0.065) & (0.064) \\ 
  & & \\ 
 C(software\_shared) & 0.175^{**} & 0.155^{**} \\ 
  & (0.068) & (0.067) \\ 
  & & \\ 
 C(data\_used) & -0.002 & 0.008 \\ 
  & (0.155) & (0.152) \\ 
  & & \\ 
 C(data\_created) & 0.061^{*} & 0.053 \\ 
  & (0.033) & (0.033) \\ 
  & & \\ 
 C(data\_shared) & 0.152^{***} & 0.152^{***} \\ 
  & (0.047) & (0.046) \\ 
  & & \\ 
 C(bso\_classification)Biology (fond.) & 0.042 & 0.045 \\ 
  & (0.037) & (0.037) \\ 
  & & \\ 
 C(bso\_classification)Chemistry & -0.048 & -0.057 \\ 
  & (0.184) & (0.180) \\ 
  & & \\ 
 C(bso\_classification)Computer and  information sciences & 0.161 & 0.125 \\ 
  & (0.110) & (0.107) \\ 
  & & \\ 
 C(bso\_classification)Earth, Ecology, Energy and applied biology & 0.011 & -0.026 \\ 
  & (0.068) & (0.067) \\ 
  & & \\ 
 C(bso\_classification)Engineering & -0.139 & -0.096 \\ 
  & (0.111) & (0.109) \\ 
  & & \\ 
 C(bso\_classification)Humanities & 0.036 & 0.016 \\ 
  & (0.080) & (0.078) \\ 
  & & \\ 
 C(bso\_classification)Mathematics & 0.009 & 0.024 \\ 
  & (0.166) & (0.163) \\ 
  & & \\ 
 C(bso\_classification)Physical sciences, Astronomy & 0.153 & 0.199 \\ 
  & (0.143) & (0.140) \\ 
  & & \\ 
 C(bso\_classification)Social sciences & 0.201^{*} & 0.174 \\ 
  & (0.121) & (0.119) \\ 
  & & \\ 
 Constant & 911.482^{***} & 928.997^{***} \\ 
  & (39.570) & (38.807) \\ 
  & & \\ 
\hline \\[-1.8ex] 
Observations & \multicolumn{1}{c}{2,812} & \multicolumn{1}{c}{2,812} \\ 
R$^{2}$ & \multicolumn{1}{c}{0.271} &  \\ 
Adjusted R$^{2}$ & \multicolumn{1}{c}{0.266} &  \\ 
Residual Std. Error (df = 2791) & \multicolumn{1}{c}{0.833} & \multicolumn{1}{c}{0.741} \\ 
F Statistic & \multicolumn{1}{c}{51.954$^{***}$ (df = 20; 2791)} &  \\ 
\hline 
\hline \\[-1.8ex] 
\textit{Note:}  & \multicolumn{2}{r}{$^{*}$p$<$0.1; $^{**}$p$<$0.05; $^{***}$p$<$0.01} \\ 
\end{longtable} 


\printbibliography

@misc{alperin_analysis_2024,
  title = {An Analysis of the Suitability of {{OpenAlex}} for Bibliometric Analyses},
  author = {Alperin, Juan Pablo and Portenoy, Jason and Demes, Kyle and Larivi{\`e}re, Vincent and Haustein, Stefanie},
  year = {2024},
  publisher = {arXiv},
  doi = {10.48550/ARXIV.2404.17663},
  urldate = {2025-05-12},
  copyright = {Creative Commons Attribution 4.0 International}
}

@unpublished{bassinet_largescale_2023,
  title = {Large-Scale {{Machine-Learning}} Analysis of Scientific {{PDF}} for Monitoring the Production and the Openness of Research Data and Software in {{France}}},
  author = {Bassinet, Aricia and Bracco, Laetitia and L'H{\^o}te, Anne and Jeangirard, Eric and Lopez, Patrice and Romary, Laurent},
  year = {2023},
  urldate = {2025-05-08}
}

@article{colavizza_analysis_2024,
  title = {An Analysis of the Effects of Sharing Research Data, Code, and Preprints on Citations},
  author = {Colavizza, Giovanni and Cadwallader, Lauren and LaFlamme, Marcel and Dozot, Gr{\'e}gory and Lecorney, St{\'e}phane and Rappo, Daniel and Hrynaszkiewicz, Iain},
  editor = {Tang, Yongli},
  year = {2024},
  month = oct,
  journal = {PLOS ONE},
  volume = {19},
  number = {10},
  pages = {e0311493},
  issn = {1932-6203},
  doi = {10.1371/journal.pone.0311493},
  urldate = {2025-05-08},
  langid = {english}
}

@article{cole_societal_2024a,
  title = {The {{Societal Impact}} of {{Open Science}}: {{A Scoping Review}}},
  shorttitle = {The Societal Impact of {{Open Science}}},
  author = {Cole, Nicki Lisa and Kormann, Eva and Klebel, Thomas and Apartis, Simon and {Ross-Hellauer}, Tony},
  year = {2024},
  month = jun,
  journal = {Royal Society Open Science},
  volume = {11},
  number = {6},
  pages = {240286},
  issn = {2054-5703},
  doi = {10.1098/rsos.240286},
  urldate = {2025-05-08},
  langid = {english}
}

@article{fu_releasing_2019a,
  title = {Releasing a Preprint Is Associated with More Attention and Citations for the Peer-Reviewed Article},
  author = {Fu, Darwin Y and Hughey, Jacob J},
  year = {2019},
  month = dec,
  journal = {eLife},
  volume = {8},
  pages = {e52646},
  issn = {2050-084X},
  doi = {10.7554/eLife.52646},
  urldate = {2025-05-12},
  copyright = {http://creativecommons.org/licenses/by/4.0/},
  langid = {english}
}

@article{hrynaszkiewicz_plos_2022a,
  title = {{{PLOS Open Science Indicators}} Principles and Definitions},
  author = {Hrynaszkiewicz, Iain and Kiermer, Veronique},
  year = {2022},
  pages = {194697 Bytes},
  doi = {10.6084/M9.FIGSHARE.21640889.V1},
  urldate = {2025-05-08},
  copyright = {Creative Commons Attribution 4.0 International}
}

@article{klebel_academic_2025,
  title = {The {{Academic Impact}} of {{Open Science}}: {{A Scoping Review}}},
  shorttitle = {The Academic Impact of {{Open Science}}},
  author = {Klebel, Thomas and Traag, Vincent and Grypari, Ioanna and Stoy, Lennart and {Ross-Hellauer}, Tony},
  year = {2025},
  month = mar,
  journal = {Royal Society Open Science},
  volume = {12},
  number = {3},
  pages = {241248},
  issn = {2054-5703},
  doi = {10.1098/rsos.241248},
  urldate = {2025-05-08},
  langid = {english}
}

@techreport{opensciencemonitoringinitiative_principles_2024,
  title = {Principles of {{Open Science Monitoring}}},
  author = {{Open Science Monitoring Initiative}},
  year = {2024},
  institution = {Minist{\`e}re de l'enseignement sup{\'e}rieur et de la recherche},
  doi = {10.52949/49},
  urldate = {2025-05-08}
}

@misc{publiclibraryofscience_plos_2023a,
  title = {{{PLOS Open Science Indicators}}},
  author = {{Public Library Of Science}},
  year = {2023},
  publisher = {Public Library of Science},
  doi = {10.6084/M9.FIGSHARE.21687686.V5},
  urldate = {2025-05-08},
  copyright = {Creative Commons Attribution 4.0 International}
}

@article{rafols_monitoring_2024,
  title = {Monitoring {{Open Science}} as Transformative Change: {{Towards}} a Systemic Framework},
  shorttitle = {Monitoring {{Open Science}} as Transformative Change},
  author = {Rafols, Ismael and Meijer, Ingeborg and {Molas-Gallart}, Jordi},
  year = {2024},
  month = apr,
  journal = {F1000Research},
  volume = {13},
  pages = {320},
  issn = {2046-1402},
  doi = {10.12688/f1000research.148290.1},
  urldate = {2025-05-08},
  langid = {english}
}

@misc{tsipouri_economic_2025a,
  title = {The {{Economic Impact}} of {{Open Science}}: {{A Scoping Review}}},
  shorttitle = {The {{Economic Impact}} of {{Open Science}}},
  author = {Tsipouri, Lena and Liarti, Sofia and Vignetti, Silvia and {Martins-Grapengiesser}, Izabella},
  year = {2025},
  month = feb,
  doi = {10.31222/osf.io/kqse5_v1},
  urldate = {2025-05-11},
  copyright = {https://creativecommons.org/licenses/by-nd/4.0/legalcode}
}

@techreport{unesco_unesco_2021a,
  title = {{{UNESCO Recommendation}} on {{Open Science}}},
  author = {{UNESCO}},
  year = {2021},
  institution = {UNESCO},
  doi = {10.54677/MNMH8546},
  urldate = {2025-05-08}
}

@article{vandewalle_code_2012a,
  title = {Code {{Sharing Is Associated}} with {{Research Impact}} in {{Image Processing}}},
  author = {Vandewalle, Patrick},
  year = {2012},
  month = jul,
  journal = {Computing in Science \& Engineering},
  volume = {14},
  number = {4},
  pages = {42--47},
  issn = {1521-9615},
  doi = {10.1109/MCSE.2012.63},
  urldate = {2025-05-12},
  copyright = {https://ieeexplore.ieee.org/Xplorehelp/downloads/license-information/IEEE.html}
}

@article{colavizza_citation_2020,
  title = {The Citation Advantage of Linking Publications to Research Data},
  author = {Colavizza, Giovanni and Hrynaszkiewicz, Iain and Staden, Isla and Whitaker, Kirstie and McGillivray, Barbara},
  editor = {Wicherts, Jelte M.},
  year = {2020},
  month = apr,
  journal = {PLOS ONE},
  volume = {15},
  number = {4},
  pages = {e0230416},
  issn = {1932-6203},
  doi = {10.1371/journal.pone.0230416},
  urldate = {2023-12-15},
  langid = {english}
}

@article{levchenko_enabling_2024,
  title = {Enabling Preprint Discovery, Evaluation, and Analysis with {{Europe PMC}}},
  author = {Levchenko, Mariia and Parkin, Michael and McEntyre, Johanna and Harrison, Melissa},
  editor = {Naudet, Florian},
  year = {2024},
  month = sep,
  journal = {PLOS ONE},
  volume = {19},
  number = {9},
  pages = {e0303005},
  issn = {1932-6203},
  doi = {10.1371/journal.pone.0303005},
  urldate = {2025-05-12},
  langid = {english}
}

@inproceedings{jeangirard_monitoring_2019,
  title = {Monitoring {{Open Access}} at a National Level: {{French}} Case Study},
  shorttitle = {Monitoring {{Open Access}} at a National Level},
  booktitle = {{{ELPUB}} 2019 23d {{International Conference}} on {{Electronic Publishing}}},
  author = {Jeangirard, Eric},
  year = {2019},
  month = jun,
  publisher = {OpenEdition Press},
  doi = {10.4000/proceedings.elpub.2019.20},
  urldate = {2025-05-12}
}

@article{langham-putrow_open_2021a,
  title = {Is the Open Access Citation Advantage Real? {{A}} Systematic Review of the Citation of Open Access and Subscription-Based Articles},
  shorttitle = {Is the Open Access Citation Advantage Real?},
  author = {{Langham-Putrow}, Allison and Bakker, Caitlin and Riegelman, Amy},
  editor = {Lozano, Sergi},
  year = {2021},
  month = jun,
  journal = {PLOS ONE},
  volume = {16},
  number = {6},
  pages = {e0253129},
  issn = {1932-6203},
  doi = {10.1371/journal.pone.0253129},
  urldate = {2025-06-09},
  langid = {english}
}

@misc{tkaczyk_crossref_2023,
  title = {Crossref Relationships between Preprints and Journal Articles},
  author = {Tkaczyk, Dominika},
  year = {2023},
  month = nov,
  publisher = {Zenodo},
  doi = {10.5281/ZENODO.10144857},
  urldate = {2025-06-09},
  copyright = {Creative Commons Zero v1.0 Universal}
}

@misc{colavizza_study_2025,
  title = {A Study on the Citation Impact of {{Open Science Indicators}} in the {{French Open Science Monitor}}},
  author = {Colavizza, Giovanni and Cadwallader, Lauren and Hrynaszkiewicz, Iain},
  year = {2025},
  publisher = {figshare},
  doi = {10.6084/M9.FIGSHARE.27822663.V2},
  urldate = {2025-06-10},
  copyright = {Creative Commons Attribution 4.0 International}
}

@article {10.7554/eLife.52646,
article_type = {journal},
title = {Meta-Research: Releasing a preprint is associated with more attention and citations for the peer-reviewed article},
author = {Fu, Darwin Y and Hughey, Jacob J},
editor = {Rodgers, Peter and Amaral, Olavo},
volume = 8,
year = 2019,
month = {dec},
pub_date = {2019-12-06},
pages = {e52646},
citation = {eLife 2019;8:e52646},
doi = {10.7554/eLife.52646},
url = {https://doi.org/10.7554/eLife.52646},
journal = {eLife},
issn = {2050-084X},
publisher = {eLife Sciences Publications, Ltd},
}

@article{10.1162/qss_a_00179,
    author = {Chaignon, Lauranne and Egret, Daniel},
    title = {Identifying scientific publications countrywide and measuring their open access: The case of the French Open Science Barometer (BSO)},
    journal = {Quantitative Science Studies},
    volume = {3},
    number = {1},
    pages = {18-36},
    year = {2022},
    month = {04},
    issn = {2641-3337},
    doi = {10.1162/qss_a_00179},
    url = {https://doi.org/10.1162/qss_a_00179},
    eprint = {https://direct.mit.edu/qss/article-pdf/3/1/18/2008278/qss_a_00179.pdf},
}

@article{fraser_relationship_2020a,
  title = {The Relationship between {{bioRxiv}} Preprints, Citations and Altmetrics},
  author = {Fraser, Nicholas and Momeni, Fakhri and Mayr, Philipp and Peters, Isabella},
  year = 2020,
  month = jun,
  journal = {Quantitative Science Studies},
  volume = {1},
  number = {2},
  pages = {618--638},
  issn = {2641-3337},
  doi = {10.1162/qss_a_00043},
  urldate = {2026-04-24},
  langid = {english}
}

@misc{henneken_linking_2011a,
  title = {Linking to {{Data}} - {{Effect}} on {{Citation Rates}} in {{Astronomy}}},
  author = {Henneken, Edwin A. and Accomazzi, Alberto},
  year = 2011,
  month = nov,
  number = {arXiv:1111.3618},
  eprint = {1111.3618},
  primaryclass = {cs},
  publisher = {arXiv},
  doi = {10.48550/arXiv.1111.3618},
  urldate = {2026-04-24},
  archiveprefix = {arXiv}
}

@misc{kucharsky_code_2020a,
  title = {Code {{Sharing}} in {{Psychological Methods}} and {{Statistics}}: {{An Overview}} and {{Associations}} with {{Conventional}} and {{Alternative Research Metrics}}},
  shorttitle = {Code {{Sharing}} in {{Psychological Methods}} and {{Statistics}}},
  author = {Kucharsk{\'y}, {\v S}imon and Houtkoop, Bobby Lee and Visser, Ingmar},
  year = 2020,
  month = feb,
  publisher = {Open Science Framework},
  doi = {10.31219/osf.io/daews},
  urldate = {2026-04-24},
  archiveprefix = {Open Science Framework},
  copyright = {https://creativecommons.org/licenses/by/4.0/legalcode}
}

@article{piwowar_data_2013a,
  title = {Data Reuse and the Open Data Citation Advantage},
  author = {Piwowar, Heather A. and Vision, Todd J.},
  year = 2013,
  month = oct,
  journal = {PeerJ},
  volume = {1},
  pages = {e175},
  issn = {2167-8359},
  doi = {10.7717/peerj.175},
  urldate = {2026-04-24},
  langid = {english}
}

\end{document}